\documentclass[aps,prd,twocolumn,showpacs,amsmath,amssymb]{revtex4}
\usepackage{graphicx,slashbox}
\usepackage{dcolumn}
\usepackage{booktabs}
\usepackage{bm,mathrsfs,bbm,amscd,epsfig}

\begin{document}

\title{A Primary Study of Heavy Baryons $\Lambda_Q$, $\Sigma_Q$, $\Xi_Q$ and $\Omega_Q$}

\author{ZHAO
Qiao-Yan$^{1}$} \thanks{E-mail:~15904816097@163.com}
\author{ZHANG Dan$^{1}$} \thanks{E-mail:~zhangdan77325@163.com}
\author{ZHANG Qiu-Yang $^{2}$}
\affiliation{\small $^{1}$School of Physical Science and Technology,
Inner Mongolia University, Hohhot 010021, P.R. China\\
$^{2}$Graduate School of Zhejiang Gongshang University, Hangzhou
310018, P.R. China}

\begin{abstract}
We perform a preliminary study of the $\frac{1}{2}^{+}$ and
$\frac{3}{2}^{+}$ ground-state baryons containing a heavy quark in
the framework of the chiral SU(3) quark model. By using the calculus
of variations, masses of $\Lambda_Q$, $\Sigma_Q$, $\Xi_Q$,
$\Omega_Q$, $\Sigma_Q^*$,
 $\Xi_Q^*$ and $\Omega_Q^*$, where $Q$ means $c$ or $b$ quark, are calculated.
With taking reasonable model parameters, the numerical results of
established heavy baryons are generally in agreement with the
available experimental data, except that those of $\Xi_Q$ are
somewhat heavier. For $\Omega_b$ with undetermined experimental mass
and unobserved $\Xi_b^*,~\Omega_b^*$, reasonable theoretical
predictions are obtained. Interactions inside baryons are also
discussed.

\end{abstract}


\maketitle

Baryons containing heavy quarks have always been interesting. In the
last decade a significant progress was achieved in the experimental
and theoretical studies of heavy hadrons. In particular, the
spectroscopy of baryons containing a singly heavy quark has obtained
special attention, mainly due to the recent experimental discoveries
\cite{ros07}. In these baryons, a heavy quark can be used as a
'flavor tag' to help us to go further in understanding the
nonperturbative QCD rather than doing the light baryons
\cite{rob08}. On the other hand, heavy baryons provide a laboratory
to study the dynamics of light quarks in the environment of heavy
quarks, such as their chiral symmetry \cite{cheng07}.

Up to date, the $\frac{1}{2}^{+}$ antitriplet charmed baryon states
($\Lambda_c^+$ \cite{la1}, $\Xi_c^+,~\Xi_c^0$ \cite{xi1}), the
$\frac{1}{2}^{+}$ and $\frac{3}{2}^{+}$ sextet charmed baryon states
[($\Omega_c$ \cite{om1}, $\Sigma_c$ \cite{si1}, $\Xi_c^{\prime}$
\cite{xip}) and ($\Omega_c^*$ \cite{om3}, $\Sigma_c^*$ \cite{si3},
$\Xi_c^*$ \cite{xi3})] have been established, while for $S$-wave
bottom baryons, only the $\Lambda_b$ \cite{la2}, $\Sigma_b,~
\Sigma_b^*$ \cite{sib}, $\Xi_b$ \cite{xib} and $\Omega_b$
\cite{omb1, omb2} have been observed \cite{pdg10}. Accordingly, a
large number of theoretical investigations have been carried out by
kinds of QCD-inspired models or methods to study masses of the
observed and expected heavy baryons. Such as various quark models
\cite{rob08,cap86,ebe05,val08}, QCD sum rules
 \cite{zhu08,wang,zhangjr}, lattice QCD \cite{mat02,liu10}, bag
model \cite{and09} and so on.

The chiral SU(3) quark model is a useful non-perturbative
theoretical tool in studying the light hadron physics. It has been
quite successful in reproducing the energies of the baryon ground
states, the binding energy of the deuteron, the nucleon-nucleon
($NN$), hyperon-nucleon ($YN$), kaon-nucleon ($KN$) and
anti-kaon-nucleon ($\bar{K} N$) scattering processes
\cite{zyzhang97,huang}. The valuable information has also been
obtained from many works on strong interactions and multiquark
clusters in this model \cite{ref-su3}. Recently it has been extended
to study the states including heavy quarks
\cite{tetra-zhang,heavy,wang10}, and provided interesting results.
All these successes inspire us to investigate masses of baryons with
one heavy quark following the above approaches. First, we briefly
introduce the framework of the chiral SU(3) quark model, which
includes the Hamiltonian and model parameters. Then the calculated
masses of $\frac{1}{2}^{+}$ and $\frac{3}{2}^{+}$ ground-state heavy
baryons involving a heavy quark are shown and discussed.

The  chiral SU(3) quark model  has been widely described in the
literature \cite{zyzhang97,huang,ref-su3,tetra-zhang,heavy,wang10}
and we recommend the reader to obtain details from those references.
Here we just give the salient feature of this model. The total
Hamiltonian of the heavy baryon containing one heavy quark ($Qqq$)
can be written as
\begin{eqnarray}
\label{hami6q} H=\sum_{i=1}^3 T_{i}-T_{G}+ V_{qq}+ \sum V_{Q q},
\end{eqnarray}
where $T_i$ is the kinetic energy operator for a single quark, and
$T_G$ is that for the center-of-mass motion. $V_{qq}$ represents the
interaction between two light quarks ($qq$).
\begin{eqnarray}
V_{qq}= V^{OGE}_{qq}+ V^{conf}_{qq} + V^{ch}_{qq},
\end{eqnarray}
where $V^{OGE}_{qq}$ is the one-gluon-exchange (OGE) interaction,
which governs the short-range perturbative QCD behavior.
$V^{conf}_{qq}$ is the confinement potential, which provides the
non-perturbative QCD effect in the long distance, taken as the
linear form in this work. $V^{ch}_{qq}$ represents the chiral fields
induced effective quark-quark potential, and describes the
non-perturbative QCD effect of the low-momentum medium-distance
range. In the chiral SU(3) quark model, it includes the scalar boson
and the pseudoscalar boson exchanges,
\begin{eqnarray}
V^{ch}_{qq} = \sum_{a=0}^8 V_{\sigma_a}+\sum_{a=0}^8 V_{\pi_a}.
\end{eqnarray}
Here $\sigma_{0},...,\sigma_{8}$ are the scalar nonet fields, and
$\pi_{0},..,\pi_{8}$ are the pseudoscalar nonet fields. The detailed
expressions of every parts can be found in Refs.
\cite{zyzhang97,huang,ref-su3,tetra-zhang,heavy,wang10}.

$V_{Qq}$ in Eq.(\ref{hami6q}) is the interaction between heavy and
light quark pairs ($Q q$) ,
\begin{eqnarray}
V_{Q q}=V_{Q q}^{OGE}+V_{Q q}^{conf}.
\end{eqnarray}
$V_{Q q}^{OGE}$ and $V_{Q q}^{conf}$ have the same forms as those of
light quark pairs. Note that following the previous works
\cite{tetra-zhang, heavy,wang10}, for $Q q$ pairs, the Goldstone
boson exchanges will not be considered in a primary study.

\begin{widetext}
{\small
\begin{center}
\begin{table}[htb]
\caption{ Model parameters for light quarks. The meson masses and
the cutoff masses: $m_{\sigma'}=980$ MeV, $m_{\kappa}=980$ MeV,
$m_{\epsilon}=980$ MeV, $m_{\pi}=138$ MeV, $m_K=495$ MeV,
$m_{\eta}=549$ MeV, $m_{\eta'}=957$ MeV, and $\Lambda=1100$ MeV for
all mesons.}\label{para}
\begin{center}
\begin{tabular}{cccccccccccc}
\hline
 $m_u$& $m_s$ & $m_\sigma$ & $g_u$ & $g_s$ & $g_{ch}$&$a_{uu}$ & $a_{us}$ &  $a_{ss}$ &$a^{0}_{uu}$& $a^{0}_{us}$&  $a^{0}_{ss}$   \\
  ~(MeV)~& ~(MeV)~ & ~(MeV)~ &  &  & &~(MeV/fm)~& ~(MeV/fm)~ &~ (MeV/fm)~ &~(MeV)~ & ~(MeV)~& ~(MeV)~ \\
\hline
313 & 470 &  595 & ~0.886~ & ~0.917~ & ~2.621~ &90.4 & 104.2 & 155.3 & $-$79.6 & $-$76.1 & -87.6  \\
\hline
\end{tabular}
\end{center}
\end{table}
\end{center}}
\end{widetext}

\begin{widetext}
 {\small
\begin{center}
\begin{table}[htb]
\caption{ Model parameters for heavy quarks.}\label{parah1}
\begin{center}
\begin{tabular}{cccccccccccc}
\hline
$g_c~~$ & $m_c$ & $a_{cu}$ & $a_{cs}$ & $a^{0}_{cu}$ & $a^{0}_{cs}$&$g_b$ & $m_b$ & $a_{bu}$ & $a_{bs}$ & $a^{0}_{bu}$ & $a^{0}_{bs}$ \\
  & ~(MeV)~~&~(MeV/fm)~~& ~(MeV/fm)~~ &~(MeV) ~~&  ~(MeV)&  & (MeV)&(MeV/fm)& (MeV/fm) &(MeV) &  (MeV)\\
\hline
0.53 &  1430 & 310.5 & 278.3 &-186.0 &  -137.9 &0.50 & 4720 & 352.6 & 310.5 &-190.0 &  -134.0  \\
     &  1550 & 339.4 & 301.5 &-223.5 &  -170.2 &    & 5100 & 391.2 & 339.9 &-281.0 &  -218.0 \\
     &  1870 & 420.6 & 368.8 &-324.2 &  -258.2 &     & 5259 & 402.4 & 352.0 &-316.4 &  -253.0 \\
\hline
0.58 & 1430 & 276.0 & 240.7 &-163.0 &  -114.3 &0.52 & 4720  & 338.0& 292.9 &-181.0 &  -124.0 \\
     & 1550 & 303.0 & 262.7 &-200.0 &  -146.5 &     & 5100  & 370.3& 323.9 &-269.0 &  -209.0 \\
     & 1870 & 376.9 & 325.0 &-298.3 &  -233.3 &    & 5259  & 390.2& 332.6 &-309.0 &  -242.5 \\
\hline
 0.60 & 1430 & 264.5 & 288.0&-155.0 &  -105.9 &0.60& 4720 & 286.5 & 236.9 &-148.0 &  -90.0 \\
     & 1550 & 290.8 & 248.9 &-191.8 &  -137.7&    & 5100 & 313.9 & 260.1 &-234.1 &  -172.0 \\
     & 1870 & 362.2 & 308.5 &-289.2 &  -223.7&     & 5259 & 325.2 & 273.6 &-270.0 &  -208.0 \\
\hline
\end{tabular}
\end{center}
\end{table}
\end{center}}
\end{widetext}

The model parameters for light quarks are taken from the previous
work \cite{zyzhang97}, which can give a satisfactory description of
the energies of the baryon ground states, the binding energy of
deuteron, the $NN$ scattering phase shifts, and the $NY$ cross
sections. As shown in Table \ref{para}, the up (down) quark mass
$m_{u(d)}$ and the strange quark mass $m_s$ are taken to be the
usual values: $m_{u(d)}=313$ MeV and $m_s=470$ MeV. The coupling
constant for scalar and pseudoscalar chiral field coupling
($g_{ch}$) is determined according to the relation
\begin{eqnarray}
\frac{g^{2}_{ch}}{4\pi} =  \frac{9}{25} \frac{g^{2}_{NN\pi}}{4\pi}
\frac{m^{2}_{u}}{M^{2}_{N}},
\end{eqnarray}
with empirical value $g^{2}_{NN\pi}/4\pi=13.67$. The masses of
mesons are taken to be the experimental values, except the $\sigma$
meson. The $m_\sigma$ is adjusted to fit the binding energy of the
deuteron. The cutoff radius $\Lambda^{-1}$ is taken to be the value
close to the chiral symmetry breaking scale \cite{ito90}. The OGE
coupling constants $g_{u}$, $g_{s}$ and the confinement strengths
$a_{qq'},a^0_{qq'}$ can be derived from the masses of ground state
baryons.

To investigate the heavy quark mass dependence, the mass of charm
quark $m_c$ is taken as several typical values $1430$ MeV
\cite{tetra-zhang}, $1550$ MeV \cite{jv04}, $1870$ MeV \cite{bs93}.
The mass of bottom quark $m_b$ is taken as $4720$ Mev
\cite{tetra-zhang}, $5100$ Mev \cite{jv05}, $5259$ Mev \cite{bs93}.

 {\small
\begin{table}[htb]
\caption{Masses (MeV) of mesons with a heavy quark. $g_c=0.58,~
m_c=1550$ MeV, and $g_b=0.52,~m_b=5100$ MeV. Experimental data are
taken from PDG \cite{pdg10}.}\label{meson}
\begin{center}
\footnotesize
\begin{tabular}{ccccccccc}
\hline
  & $D$ & $D^*$ & $D_s$ & $D_s^*$& $B$ & $B^*$ & $B_s$ & $B_s^*$\\
\hline
Exp. & 1869.6 & 2007.0&1968.5 & 2112.3& 5279.2 & 5325.1 & 5366.3 & 5415.4 \\
Thoer. & 1869.8 & 2007.1 & 1968.6 & 2112.3& 5279.3 & 5325.1 & 5366.0 & 5415.3 \\
\hline
\end{tabular}
\end{center}
\end{table}}

To test their effects on other parameters and on the spectrum, the
OGE coupling constants for heavy quarks are taken as three values in
an estimated range \cite{tetra-zhang}, i.e. $g_{c}=0.53,~0.58,~0.60$
and $g_{b}=0.50,~0.52,~0.60$. The confinement strengths including a
heavy quark ($a_{Qq}$,~$a_{Qq}^{0}$) are determined by fitting the
masses of heavy mesons $D$, ~$D^{*}$,~ $D_{s}$,~ $D_{s}^{*}$ and
$B$, ~$B^{*}$,~ $B_{s}$,~ $B_{s}^{*}$, respectively. The parameters
about heavy quarks are tabulated in Table \ref{parah1}. The
corresponding numerical masses of heavy mesons are exactly
consistent with the experimental values. As an example, the results
with $g_c=0.58,~ m_c=1550$ MeV and $g_b=0.52,~ m_b=5100$ MeV are
listed in Table \ref{meson}.

With all parameters determined, masses of the $\frac{1}{2}^+$ lowest
lying ground-state $\Lambda_Q$, $\Sigma_Q$, $\Xi_Q$, $\Omega_Q$ and
$\frac{3}{2}^+$ $S$-wave $\Sigma_Q^*$, $\Xi_Q^*$, $\Omega_Q^*$,
where $Q$ means $c$ or $b$ quark, can be calculated by the calculus
of variations. The harmonic-oscillator width $b_u$ is taken as the
variational parameter. Compared to the experimental data, the
numerical results can be found in Table \ref{mass}, and some other
theoretical predictions are illustrated as well.
\begin{widetext}
{\small
\begin{center}
\begin{table}[htb]
\caption{Masses (MeV) of baryons with a heavy quark, accompanied by
some other theoretical predictions. Experimental data are taken from
PDG \cite{pdg10}.}\label{mass}
\begin{center}
\footnotesize
\begin{tabular}{ccccccccc|ccccccccc}
\hline
$g_c$ & $m_c$ &  $\Lambda_c$& $\Sigma_c$ &$\Xi_c$&$\Omega_c$&$\Sigma_c^*$& $\Xi_c^*$& $\Omega_c^*$
&$g_b$ & $m_b$ &  $\Lambda_b$& $\Sigma_b$ &$\Xi_b$&$\Omega_b$&$\Sigma_b^*$& $\Xi_b^*$& $\Omega_b^*$\\
\hline
 0.53& 1430 & 2307.7 & 2473.7 & 2541.9 & 2692.5& 2536.7 &2653.3 &2766.7 & 0.50&4720&5644.2&5843.2& 5879.7 &6037.0 &5865.6 &5971.8 &6083.4 \\
     & 1550 & 2306.9 & 2485.5 & 2546.5 & 2697.7& 2548.0 &2662.5 &2782.2 &     &5100&5643.8&5858.8& 5885.7 &6043.6 &5880.9 &5984.2 &6090.7\\
     & 1870 & 2305.9 & 2518.2 & 2559.4 & 2713.4& 2580.1 &2689.4 &2799.5 &     &5259&5643.6&5863.2& 5887.5 &6046.1 &5885.2 &5987.9 &6093.8  \\
\hline
0.58 & 1430 & 2308.8 & 2457.9 & 2535.9 & 2684.1 & 2521.5 &2640.1 &2767.6 &0.52 &4720 &5644.7& 5837.0& 5877.2 &6031.1 &5859.5 &5966.4 &6079.6 \\
     & 1550 & 2307.9 & 2468.6 & 2539.8 & 2688.5 & 2532.0 &2648.6 &2772.5 &     &5100 &5644.1& 5849.8& 5882.1 &6039.7 &5867.7 &5975.1 &6086.8 \\
     & 1870 & 2306.0 & 2497.8 & 2551.2 & 2702.4 & 2560.6 &2672.5 &2787.9 &     &5259 &5643.9& 5857.7& 5885.1 &6041.9 &5880.0 &5982.8 &6088.9   \\
\hline
0.60 & 1430 & 2309.3 & 2452.3 & 2533.8 & 2681.3& 2516.8 &2635.8 & 2765.0 & 0.60&4720 &5646.2& 5815.0& 5868.8 &6023.3 &5838.3 &5948.5 &6068.1 \\
     & 1550 & 2308.4 & 2462.6 & 2537.6 & 2685.6& 2526.7 &2643.9 & 2769.6 &     & 5100&5645.4& 5858.8& 5872.6 &6027.4 &5848.7 &5956.8 &6072.8 \\
     & 1870 & 2306.5 & 2490.8 & 2548.2 & 2698.4& 2554.2 &2666.9  & 2783.7&     & 5259&5645.4& 5858.8& 5872.6 &6030.0 &5852.9 &5960.8 &6075.8  \\
\hline
Exp. &   & 2286.5 & 2453.8 & 2471.0 &2697.5 & 2518.0 &2646.6 &2768.3 & & & 5620.2 & 5807.8& 5792.4 &   &5829.0 &   &  \\
\hline
Ref. \cite{ebe05}&   & 2297 & 2439 & 2481 & 2698 & 2518 & 2654 & 2768 & & & 5622 & 5805 & 5812 & 6065 & 5834 & 5963& 6088  \\
Ref. \cite{zhu08}&   & 2271 & 2411 & 2432 & 2657 & 2534 & 2634 & 2790 & & & 5637 & 5809 & 5780 & 6036 & 5835 & 5929& 6063  \\
Ref. \cite{wang}&(GeV)& 2.26 & 2.40 & 2.44& 2.70 & 2.48 & 2.65 & 2.79 & & & 5.65 & 5.80 & 5.73 & 6.11 & 5.85 & 6.02& 6.17  \\
Ref. \cite{zhangjr}&(GeV)& 2.31& 2.40& 2.48& 2.62& 2.56 & 2.64 & 2.74 & & & 5.69 & 5.73 & 5.75 & 5.89 & 5.81 & 5.94& 6.00  \\
Ref. \cite{mat02}&   & 2290 & 2452 & 2473 & 2678 & 2538 & 2680 & 2752 & & & 5672 & 5847 & 5788 & 6040 & 5871 & 5959& 6060  \\
\hline
$\diamondsuit$~0.58 & 1550  & 2295.1 & 2452.8 & 2525.4 & 2672.8 & 2503.6& 2629.8 & 2752.5 &$\diamondsuit$~0.52~~ &5100 & 5651.0 & 5806.8 & 5868.3 & 6021.1 & 5823.9 & 5933.1& 6054.4  \\
\hline
\end{tabular}
\end{center}
\begin{flushleft}
{Note that $J^P$ of baryons with '*' are $\frac{3}{2}^+$, and others
are $\frac{1}{2}^+$. In addition, the last line '$\diamondsuit$'
lists the masses (MeV) of heavy hadrons with $g_c=0.58,~m_c=1550$
MeV, and $g_b=0.52,~m_b=5100$ MeV after varying the confinement
strengths.}
\end{flushleft}
\end{table}
\end{center}}
\end{widetext}

From Table \ref{mass}, we can see that for $J^P=\frac{1}{2}^+$, the
numerical values of $\Lambda_c$ are generally about $20$ MeV higher
than the experimental one. For $\Sigma_c$, the largest difference is
$64$ MeV ($2518.2$ MeV$-2453.8$ MeV), while the closest mass
($2452.3$ MeV) is obtained with $g_c=0.6$ and $m_c=1430$. The
results of $\Xi_c$ are somewhat poor, which are about $62\sim 88$
MeV higher. The results of $\Omega_c$ are at most $16$ MeV far from
the observed value, but the exact mass ($2697.7$ MeV) appears when
$g_c=0.5$ and $m_c=1550$. Predictions of baryons with $b$ quark are
about $25$ MeV heavier for $\Lambda_b$, and $76\sim95$ MeV higher
for $\Xi_b$. The nearest value of $\Sigma_b$ ($5815.0$MeV) can be
found when $g_b=0.6$ and $m_b=4720$, and others are $29\sim55$ MeV
heavier. For $\Omega_b$ with uncertain experimental data, our
average results are compatible with the observed value ($6054.4$
MeV) from Ref. \cite{omb2} and the theoretical predictions from
Refs. \cite{mat02, zhu08}. When $J^P=\frac{3}{2}^+$ the situation
has been improved. The calculated values which are consistent with
the experimental ones are obtained by $g_c=0.6$ and $m_c=1430$ for
$\Sigma_c^*$ ($2516.6$ MeV),
$g_c=0.58$ and $m_c=1550$ for $\Xi_c^*$ ($2648.6$ MeV),
$g_c=0.6$ and $m_c=1550$ for $\Omega_c^*$ ($2769.6$ MeV), $g_b=0.6$
and $m_b=4720$ for $\Sigma_b^*$ ($5838.3$ MeV). For unobserved
$\Xi_b^*$ and $\Omega_b^*$, our predictions are similar to those
from Ref. \cite{ebe05}.  It is worth noting that Ref. \cite{wang10}
give predictions of $\Lambda_c$ ($M_{\Lambda_c}=2269$ MeV) and
$\Sigma_c$ ($M_{\Sigma_c}=2436$ MeV) using the same model as ours by
taking $g_c=0.53$ and $m_c=1550$MeV. While compared with our present
results, their corresponding meson masses $m_D=1883$ MeV,
$m_{D^*}=1947$ MeV become far away from the experimental values.

To lower the calculated values, we vary the confinement parameters
in a reasonable range, which means that masses of heavy mesons are
roughly consistent with the experimental data. When confinement
strengths are changed, the results of $\Sigma_Q$ are in good
agreement with the observed ones, and masses of $\Xi_Q$ are
decreased by $10\sim20$ MeV. Most of $\Lambda_c$ masses are also
decreased, and the smallest one ($2292.6$ MeV) is only $6$ MeV
higher. However, little action is played for the calculation of
$\Lambda_b$. In general, the numerical values are reduced less than
$20$ MeV for $\Omega_Q$, $60$ MeV for $\Sigma_Q^*$, $50$ MeV for
$\Xi_Q^*$, and $30$ MeV for $\Omega_Q^*$, which leads to the
predictions being $10\sim30$ MeV lower than the available observed
ones. It should be noted that after changing the parameters,
compared with the experimental data, the corresponding calculated
masses of some heavy mesons are about $1\%$ shifted. This relatively
small difference can be acceptable in theory. As an example, with
$g_c=0.58,~m_c=1550$ MeV, and $g_b=0.52,~m_b=5100$ MeV, the changed
confinement parameters and corresponding masses of mesons are listed
in Table \ref{mass2}, and those of baryons are illustrated in the
last line with '$\diamondsuit$' of Table \ref{mass}.

{\small
\begin{table}[htb]
\caption{Masses (MeV) of heavy mesons after varying confinement
strengths with $g_c=0.58,~m_c=1550$ MeV, and $g_b=0.52,~m_b=5100$
MeV. Same units of $a_{Qq}, a_{Qq}^0$ as in Table
\ref{parah1}.}\label{mass2}
\begin{center}
\begin{tabular}{cccccccc}
\hline
$a_{cu}$ & $a_{cs}$ & $a^{0}_{cu}$  & $a^{0}_{cs}$  & $a_{bu}$&$a_{bs}$  & $a^{0}_{bu}$  & $a^{0}_{bs}$ \\
295.3 &238.7 & -198.0 & -136.5 & 247.6 & 164.5& -200.8  & -130.0  \\
\hline
$D$ & $D^*$ & $D_s$ & $D_s^*$&$B$ & $B^*$ & $B_s$ & $B_s^*$\\
1859.1 &1993.7 &1964.8 &2098.1&5292.1 &5325.1 &5386.0 &5415.3\\
\hline
\end{tabular}
\end{center}
\end{table}}

{\small
\begin{table}[htb]
\caption{The effects of meson exchanges between light quark
pairs.}\label{inter}
\begin{center}
\begin{tabular}{ccccc}
\hline
$J^P$&Baryons &~ Attractions~ & ~Repulsions ~& ~No effects   \\
\hline
$\frac{1}{2}^+$~~~&$\Lambda_Q$ & $\pi, \epsilon, \sigma$ & $\eta, \eta', \sigma'$ & $K, \kappa$ \\
 & $\Sigma_Q$ & $\pi, \eta, \eta', \sigma', \epsilon, \sigma$ & $-$  & $K, \kappa$ \\
 & $\Xi_Q$ & $K, \eta, \sigma$ & $\eta', \kappa, \epsilon$ & $\pi, \sigma'$ \\
 & $\Omega_Q$ & $ \epsilon, \sigma$ & $-$ & $\pi, K, \eta, \eta', \sigma', \kappa$ \\
\hline
$\frac{3}{2}^+$~~~&$\Sigma_Q^*$ & $\pi, \eta, \eta', \sigma', \epsilon, \sigma$ & $ - $ & $ K, \kappa$ \\
 & $\Xi_Q^*$ & $K, \eta', \kappa, \sigma$ & $ \eta, \epsilon$ & $\pi, \sigma'$ \\
 & $\Omega_Q^*$ & $\eta, \eta', \epsilon, \sigma$ & $-$ & $\pi, K, \sigma', \kappa$ \\
\hline
\end{tabular}
\end{center}
\end{table}}

Next, let us turn to interactions in these heavy baryons. The
effects of meson exchanges between light quarks are shown in Table
\ref{inter}, which are only related to $g_{ch}$, the masses and the
cutoff masses of mesons. For all baryons considered here, OGE
interactions are attractive, while actions of the confinement
potentials depend on the confinement strengths. When we keep $m_Q$
unchanged, with $g_Q$ increasing, the OGE attractions will increase,
too, and accordingly the confinement attractions decrease (or
repulsions increase). This is obvious. Here unchanged $m_Q$
indicates that the total force for a heavy baryon is changeless. The
attraction of OGE grows, which certainly accompanies by that of
confinement reducing with other conditions fixed. On the other hand,
when $g_Q$ does not change, the forces of OGE almost keep
invariable. With $m_Q$ increasing, confinement attractions will
increase (or repulsions will decrease). Similarly, $m_Q$ growing
implies that the total force becomes larger for the baryon cluster,
the confinements need more attractive when other factors remain
unchanged.

In summary, we have performed a primary study on $\frac{1}{2}^+$ and
$\frac{3}{2}^+$ ground-state baryons with one heavy quark ($c$ or
$b$) in the chiral SU(3) quark model. The calculated masses of
established heavy baryons are generally in agreement with the
available experimental data, except that those of $\Xi_Q$ are
somewhat heavier. Reasonable theoretical predictions of $\Omega_b$
with uncertain experimental mass and unobserved
$\Xi_b^*,~\Omega_b^*$ are presented. Meanwhile, interactions inside
baryons are analyzed, too. It is suggested that our predictions
could serve as a useful complementary tool for the interpretation of
heavy hadron spectra. However, there are several problems in our
present study deserving further discussions, for example, the
effects of vector meson exchanges. Furthermore, we hope that the
same approach is applied to explore more properties of heavy baryons
(such as the spectra of baryons with two or three heavy quarks, or
strong interactions including heavy baryons), and test the model
parameters compared to the experimental data. All these topics will
be researched in future.

\begin{acknowledgements}
Authors would like to thank Professor Zhang Zong-Ye and Post-doctor
Wang Wen-Ling for helpful discussions. This work was supported by
the China Postdoctoral Science Foundation Funded Project
(20100471491), the Natural Science Foundation of Inner Mongolia
(2010MS0101), the Inner Mongolia Educational Foundation (NJzy08006),
SPH-IMU (Z20090143), and the Graduate Program of Inner Mongolia
University.
\end{acknowledgements}

\end{document}